\documentclass{elsart}

\usepackage{graphics}
\usepackage{graphicx}
\journal{Solid State Communications}
\usepackage{amssymb}

\begin{document}

\begin{frontmatter}



\title{Resonant tunnelling in interacting 1D systems with an AC
modulated gate}

 \author[label1]{A.Komnik} 
 \ead{komnik@uni-freiburg.de}
and
 \author[label2]{A.O.Gogolin}
 \address[label1]{Physikalisches Institut, Albert--Ludwigs--Universit\"at,
 Hermann--Herder--Str. 3, D--79104 Freiburg, Germany}
 \address[label2]{Department of Mathematics, Imperial College London, 180
 Queen's Gate, SW7 1BZ London, United Kingdom}
\author{}

\address{}

\begin{abstract}
We present an analysis of transport properties of a system consisting of two 
half-infinite interacting one-dimensional wires connected to a single
fermionic site, the energy of which is subject to a periodic time modulation. 
Using the properties of the exactly solvable Toulouse point we derive an
integral equation for the localised level Keldysh Green's function which governs 
the behaviour of the linear conductance. We investigate
this equation numerically and analytically in various limits.
The period-averaged conductance $G$ displays a surprisingly rich behaviour
depending on the parameters of the system. The most prominent
feature is the emergence of an intermediate temperature regime
at low frequencies, 
where $G$ is proportional to the {\it line width} of the
respective static conductance saturating at a non-universal 
frequency dependent value at lower temperatures.
\end{abstract}

\end{frontmatter}

\section{Introduction}
The future advances of the microelectronics depend crucially on the
developments in the field of miniaturisation of electronic circuits. The
ultimate basis devices are single molecules or even atoms supplemented by
three electrodes: source, drain and an additional gating electrode responsible
for the current flow control. Such structures can, however, be so small 
that quantum effects have to be taken into account. The simplest possible
realisation of such a device is a single fermionic level coupled to two
metallic electrodes and subjected to an electrostatic interaction
with a gate electrode. Such structures are also called single-state quantum
dots. In the ideal situation of very small lateral device dimensions the 
contacting electrodes can be considered to be one-dimensional (1D). For
instance, thin metallic single-wall carbon nanotubes (SWNTs) can play the role
of such wirings. 
However, 
the low energy sector of 1D metals cannot be described by the conventional
Fermi liquid theory, they are instead Luttinger liquids  
(LLs, for a recent review see e.g. \cite{book}). 
Generally, the transport problem of a quantum dot
coupled to  LLs is not integrable \cite{furusaki}. 
Nevertheless, in some situations there exist unitary
transformations which allow one to map the Hamiltonian of the system
onto a diagonalisable quadratic one. In such situations all 
transport properties including the noise power spectra 
are directly accessible \cite{ourPRL,ourPRB}.  

In addition to the static conductance properties future applications in 
integrated circuits demand also a detailed knowledge of the device response to
AC currents and especially to an AC modulated gating. Thus far numerous 
studies of quantum dots under such conditions have been made
\cite{meirwingreen,jauho,langreth}. 
However, all 
of them deal with quantum dot systems contacted by {\it non-interacting} 
electrodes. The purpose of this work is to gain initial insights into 
phenomena taking place in AC-modulated devices coupled to 
{\it interacting} LL electrodes.
The outline of the paper is as follows. 
We begin by summarising the results for non-interacting systems
in Section \ref{NI}, where we also elaborate on the meaning
of the low and high frequency approximations.
We analyse the dynamic conductance for LL electrodes
in Section \ref{I}.

\section{Formulation of the problem and non-interacting results}
\label{NI}
The Hamiltonian under investigation,
\begin{eqnarray}                     \label{H0}
 H = H_K + H_t + H_C \, ,
\end{eqnarray}
contains three parts. The first one, $H_K$, describes the dynamics of the
resonant level and the electrons in the right/left (R/L) electrodes (we shall
also call them leads),
\begin{eqnarray} \nonumber
H_K = \Delta(t) d^\dag d + \sum_{i=R,L} H_0[\psi_i] \, ,
\end{eqnarray}
where the dynamic gating of the resonant level is described by 
explicitly time dependent level energy $\Delta(t)$. As long
as the Kondo temperature is small, the spin effects are trivial for
resonant tunnelling problem, therefore we work with a spinless
model and briefly discuss the results for the spinful situations in Conclusions.
The tunnelling between the
resonant level and the electrodes with energy independent amplitudes
$\gamma_{R,L}$ is taken care of by the second term, 
\begin{eqnarray}          \nonumber
 H_t = \sum_i  \gamma_i[ d^\dag \psi_i(0) + \mbox{h.c.}] \, .
\end{eqnarray}
In (\ref{H0}), $H_C$ stands for the electrostatic
Coulomb interaction between the leads and the dot,
\begin{eqnarray} \nonumber
 H_C =  \lambda_C d^\dag d \, \sum_i \, \psi_i^\dag(0) \psi_i(0) \, .
\end{eqnarray}
In the non-interacting case, when $H_0[\psi_i]$ describes free fermions in 1D
and when $\lambda_C=0$ the non-linear
$I(V)$ can be shown in the static case to be given by the Meir-Wingreen 
formula \cite{meirwingreen}, 
\begin{eqnarray}                          \label{mwin}
 I(V) = -\frac{e}{\pi} \int d\omega [f_L(\omega)-f_R(\omega)] \, \frac{\Gamma_L
 \Gamma_R}{\Gamma_R + \Gamma_L} \, \mbox{Im} \, D^R(\omega) \, ,
\end{eqnarray} 
where $f_{L,R}(\omega) = n_F(\omega\mp eV/2)$ are the Fermi distribution
functions in the leads and $D^R(\omega)$ denotes the Fourier component 
of the retarded Green's function $D^R(t,t')$ of the localised level.
$\Gamma_{R,L}=\pi \rho_0 \gamma^2_{R,L}$ are the dimensionless conductances of 
the contacts ($\rho_0$ denotes the density of states in the leads). 
For the linear conductance one obtains then
\begin{eqnarray}                       \label{formulaforG}
 G/G_0 = \frac{1}{4 T} \int d \omega \cosh^{-2} (\omega/2T) T(\omega) \, 
\end{eqnarray}
where $T(\omega)=-\Gamma_L \Gamma_R \mbox{Im} \, D^R(\omega)/(\Gamma_R +
\Gamma_L)$ is the
transmission coefficient of the electrons through the double constriction and
$G_0 = e^2/h$ denotes the conductance quantum. In 
the static case $D^R(t,t')$ depends only on the 
difference $t-t'$ while in case of
the dynamic gating it depends on $t-t'$ as well as $t+t'$. In such situations
one is usually interested in the current (or conductance) averaged over the
period of the gate voltage oscillations. The non-interacting situation has
been discussed in detail in Ref.\cite{jauho}. Here
we are interested in the linear conductance rather then in the full
$I(V)$. Therefore we recapitulate and somewhat extend the results of
Ref.\cite{jauho} in order to be able to compare to the interacting results
we're presenting in the next Section. 
In this situation $D^R(t,t')$ can be solved for 
exactly leading to \cite{langreth}
\begin{eqnarray}                           \label{exNID}
 D^R(t,t') = - i \Theta(t-t') \exp\left[ -i \int_{t'}^{t} d\tau \Delta(\tau) -
 \frac{1}{2}(\Gamma_L + \Gamma_R) (t-t') \right] \, .
\end{eqnarray}
Assuming the gating is harmonic, $\Delta(t) = \Delta_0 + \Delta_1
\cos (\Omega t)$, one can easily perform the time integrations and the
averaging over the period $2\pi/\Omega$ with respect to $t+t'$. Then, using
the Bessel $J_0$ function one obtains \cite{jauho}: 
\begin{eqnarray}                                    \nonumber 
 \bar{D}^R(t-t') = -i \Theta(t-t')
 e^{[-i\Delta_0-(\Gamma_L+\Gamma_R)/2](t-t')}  J_0\left\{
 \frac{2\Delta_1}{\Omega} \sin [ \Omega(t-t')/2] \right\}\;.
\end{eqnarray}
As a result we obtain for the imaginary part [from now on we measure all
energies in units of $(\Gamma_R+\Gamma_L)/2$]
\begin{eqnarray}
 \mbox{Im} \bar{D}^R(\omega) = - \sum_{n=-\infty}^\infty \frac{
 J_n^2(\Delta_1/\Omega)}{(\omega-\Delta_0+n\Omega)^2+1} \, .
\end{eqnarray}
Insertion of this relation into (\ref{formulaforG}) 
yields then the following analytic formula for the
linear conductance
\begin{eqnarray}            \label{exNIresult}
 G/G_0=\frac{4\Gamma_R \Gamma_L}{(\Gamma_R +\Gamma_L)^2} \frac{1}{2 \pi T}
  \sum_n 
  J_n^2(\Delta_1/\Omega) \mbox{Re}\left\{ \Psi'\left[\frac{1}{2}
 +\frac{1+i(n\Omega-\Delta_0)}{2 \pi T}\right]\right\} \, .
\end{eqnarray}
At least in the non-interacting situation the asymmetry effects are trivial
and are only reflected by the corresponding prefactor, so that we restrict our
analysis to $\Gamma_R=\Gamma_L$. 
\begin{figure}
\vspace*{0.0cm}
\includegraphics[scale=0.73]{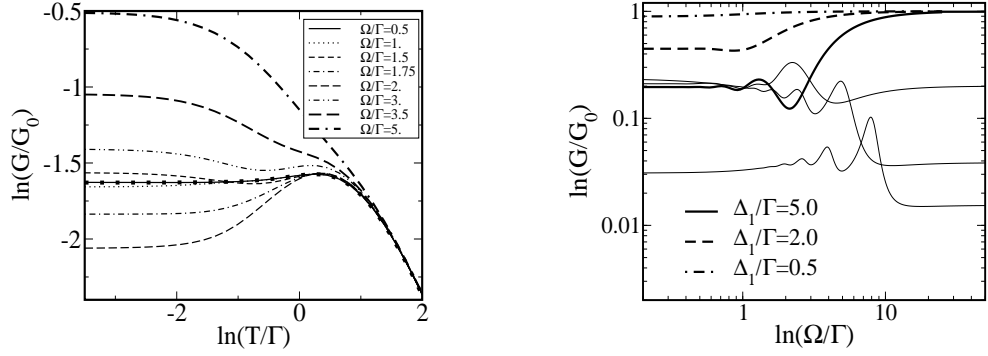}
\caption[]{\label{combpicture1} \emph{Left panel:} Temperature dependence of 
the period averaged linear
conductance for different gate oscillation frequencies and
$\Delta_1/\Gamma=5$, $\Delta_0=0$. The thick dotted line (not included into
the legend) represents the result of the adiabatic approximation. 
\emph{Right panel:} Thick lines: frequency dependence
of the asymptotic linear conductance at $T=0$ for different values of
$\Delta_1$, $\Delta_0=0$. Thin lines represent the cases of nonzero
$\Delta_0/\Gamma=2, 5, 8$ and $\Delta_1/\Gamma=5$ (from above). Notice the 
double logarithmic scale.}
\vspace*{-0.0cm}
\end{figure}
At $T\gg \Gamma$, the conductance ceases to depend on either $\Omega$ or
$\Delta_i$ ($i=1,2$), and is given by $G/G_0 \approx \pi/4T$. In the
opposite limit of low temperatures the main feature of $G$ is generic:
it saturates as $T$ approaches zero. However, the saturation values are
not the same at different oscillation frequencies. For very small 
frequencies and $\Delta_0 \lesssim \Delta_1$, $G(0)/G_0 \approx 1/\Delta_1$
whereas for $\Delta_0 \gg \Delta_1$, the conductance  
decays as $\sim \Delta_0^{-2}$, see Fig.\ref{combpicture1}. From the
physical point of view it can be understood as follows. In the static
situation  due to finite
tunnelling amplitude the localised level and the leads hybridise, so that the
spectral function of the dot state has a finite width $\sim \Gamma$. Then, as
long as  $\Delta_0 \lesssim \Delta_1$, the 
level spends a fraction $\nu = \Gamma/\Delta_1$ of the
oscillation period within `reach' of the electrodes. That's why the averaged
conductance is proportional to $\nu$. On the other hand, for $\Delta_0 \gtrsim
\Delta_1$ the overlap of two areas cannot ever become maximal. 
 That leads to a fast decay of the averaged $G$ as a function of
$\Delta_0$. In the opposite case of
very high frequencies the average time, which an electron needs to be
transferred through the constriction, $1/\Gamma$, is much larger than the
oscillation period, so that the particles only see the level being `frozen' at
its average energy $\Delta_0$, which means that $G/G_0 \approx
(1+\Delta_0^2)^{-1}$ in that limit. The crossover between large and small
frequencies is highly non-monotonous, see Fig.\ref{combpicture1}. There are
local maxima at $\Omega=\Delta_0/(2n+1)$ and minima at $\Omega=\Delta_0/2n$ 
($n$ is a natural number), which correspond to absorption of multiple energy
quanta.  

Of course, all these results can be recovered using the corresponding
asymptotic  
expansions of the exact result (\ref{exNIresult}), see Appendix. Moreover,
the asymptotic behaviour of the averaged conductance can be extracted from the
static level Green's function without knowledge of the exact dynamic one
(\ref{exNID}). 
At very low frequencies $\Omega \ll \Gamma$ the system can be regarded to be 
in equilibrium with respect to the oscillations at any time and
the obvious approximation is thus given by (\ref{mwin}) 
with the period-averaged 
transmission coefficient $T(\omega,t)=[1+(\omega-\Delta(t))^2]^{-1}$ 
(we are still dealing with
the symmetric situation $\Gamma_R=\Gamma_L$)
\begin{eqnarray}
 \bar{T}(\omega) = \frac{\Omega}{2 \pi} \int_0^{2\pi/\Omega} \, dt \,
 T(\omega,t) 
 = \mbox{Re}
 \left\{ [1+i(\omega-\Delta_0)]^2 + \Delta_1^2 \right\}^{-1/2} \, .
\end{eqnarray}
The results of this \emph{adiabatic} approximation are depicted in 
Fig.\ref{combpicture1} and
describes very well the behaviour of $G$ for frequencies up to $\Gamma$. 
We wish to stress here that $\Omega\to 0$ limit is non-trivial
and is {\it not} equivalent to simply taking $\Delta_1=0$. Since
the usual experimental values of $\Gamma$ range between 0.1 and 1 $\mu$eV, the
static approximation is expected to be valid to fairly high 
frequencies of the order of several MHz. 
The opposite limit of $\Omega\gg \Gamma$ is even simpler. In that situation
one can use the formula for the static case with
$\Delta(t)\to\Delta_0$. This approach we
refer to as \emph{anti-adiabatic} approximation.

\section{Interacting leads}\label{I}
It has been shown in a number of recent works that the resonant tunnelling 
between
correlated electrodes is completely different from the non-interacting case
\cite{kf,furusakinagaosa,ng,ourPRL,gornyi,ourPRB,huegleegger}. 
As in the non-interacting situation, we start with the Hamiltonian (\ref{H0})
with $\lambda_C \neq 0$. However, contrary to the previous case, we model the
leads by half-infinite LLs. It turns out 
that in the static case the problem can be solved exactly at the so-called 
Toulouse point, when the LL interaction parameter $g=1/2$ and $\lambda_C =
2\pi$ \cite{ourPRL}. Generally, the transport properties can be extracted in
two different ways, either using the equations of motion method \cite{ourPRL}, 
or via Keldysh diagram technique \cite{ourPRB}. The latter approach is,
however, considerably easier to apply in the case of a dynamic resonant
level. After a unitary transformation, refermionization and introduction of the
Majorana components of the new fermionic degrees of freedom, $\xi$, $\eta$,
and those for the dot level, $a$, $b$, the resulting
Hamiltonian can be written in the form $H=H_0 + H_t$, where 
(see \cite{ourPRL,ourPRB} for details of the mapping; we use the
same notation)
\begin{eqnarray}
 H_0 = i \Delta(t) ab + i \int dx [\eta(x) \partial_x \eta(x) + \xi(x)
 \partial_x \xi(x) &+& V \xi(x) \eta(x)]  \, ,
\end{eqnarray}
and $H_t = i \gamma  b \xi(0)$, while the current operator is given by
 $J = -i \gamma b \eta(0)$.
Calculating the average of this operator using the $S$ matrix
containing the $H_t$ coupling, one can derive the following expression
for the full (time-dependent) non-linear $I-V$ of the system, 
\begin{eqnarray}
 I(V) =  \frac{e}{8 \pi} \Gamma \int d \omega 
 [n_F(\omega+eV)-n_F(\omega-eV)]
\mbox{Im} D^R_{bb} (\omega,t) \, ,
\end{eqnarray}
where $\Gamma=\gamma^2$ and $D^R_{bb}(\omega,t)$ is the Fourier component of
the retarded Green's function (GF) of the $b$ Majoranas with respect to the 
difference of time arguments, 
\begin{eqnarray}
 D^R_{bb}(\omega, t=\tau+\tau') = \int d(\tau-\tau')\exp[i \omega
 (\tau-\tau')] D^R_{bb}(\tau,\tau') \, .
\end{eqnarray} 
As in the non-interacting case, the linear conductance turns
out to be given by the same formula (\ref{formulaforG}). 
The meaning of the transmission
coefficient $T(\omega)$, of course, changes. Now it is related to the
probability for the `new' fermions (see Ref.\cite{ourPRL}) to be scattered off
the resonant level. According to Ref. \cite{ourPRB} in the static case one 
obtains 
\begin{eqnarray}                        \label{interactingTomega}
T(\omega)=\omega^2/[(\omega^2-\Delta^2)^2+\omega^2] \, , 
\end{eqnarray}
while for
dynamic gating the effective $T(\omega)$ has to be found 
via solution of corresponding 
Dyson equation for the retarded GF, 
\begin{eqnarray}                   \label{inteqfull}
 D^R_{bb}(t,t') = D^{(0)R}_{bb}(t,t') + \Gamma \int d \tau d \tau'
 D^{(0)R}_{bb}(t,\tau) G^{(0)R}_{\xi \xi} (\tau-\tau') D^R_{bb}(\tau,t') \, ,
\end{eqnarray}
where $G^{(0)R}_{\xi \xi} (\tau-\tau')= -i\delta(\tau-\tau')/2$ is the retarded
GF of the lead Majoranas for $\gamma=0$ and 
\begin{eqnarray}
 D^{(0)R}_{bb}(t,t') = -i \Theta(t-t') \cos [\int^{t}_{t'} d \tau \Delta(\tau)]
\end{eqnarray}
is the bare retarded GF of the resonant level Majoranas. 
\begin{figure}
\vspace*{0.0cm}
\includegraphics[scale=0.3]{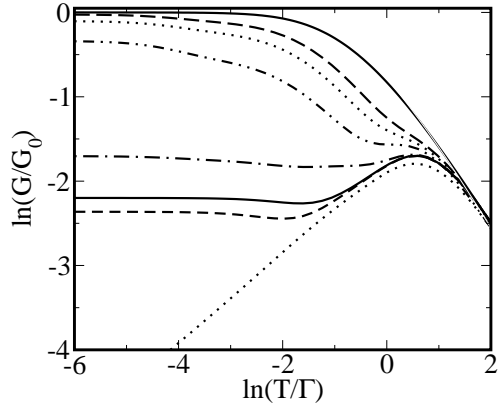}
\caption[]{\label{Ipicture1} Temperature dependence of the averaged
conductance in the interacting case for different gate frequencies $\Omega$
and $\Delta_0=0$. The uppermost curve represents the result of the
anti-adiabatic approximation 
while the lowermost curve is the result of the adiabatic approximation. The
curves in between correspond to $\Omega/\Gamma=4$, $3.5$, $3$, $1$, $0.1$ and 
$0.05$ (from above). Notice the double logarithmic scale.}
\vspace*{-0.0cm}
\end{figure}
Unfortunately we could not solve Eq.(\ref{inteqfull}) analytically.
Before we present numerical results, let us discuss
the limiting cases of low and high $\Omega$. 
As expected from the analysis of the non-interacting case, these
limiting cases should be reliably described by the adiabatic and 
the anti-adiabatic approximations, respectively. Let us
first investigate the most interesting case of 
zero off-set $\Delta_0=0$. The $\Omega
\rightarrow \infty$ limiting case is trivial and the conductance
is given by the corresponding static result with $\Delta_{0,1}=0$, 
see the uppermost curve in Fig.\ref{Ipicture1}. 
In the case of slow oscillations we resort to the adiabatic
approximation, which can be implemented in the same way as in the
non-interacting situation, namely by averaging the corresponding
static transmission coefficient (\ref{interactingTomega}) over the
oscillation period: 
\begin{eqnarray}
 \bar{T}(\omega) = \left| \frac{2 \omega}{\Delta_1^2} \mbox{Im}\left\{
 [2(\omega^2-i\omega)/\Delta_1^2-1]^2-1\right\}^{-1/2} \right| \, .
\end{eqnarray} 
Evaluation of the linear conductance using this approximation results in the
lowest curve in Fig. \ref{Ipicture1}. The main feature is that $G$ vanishes at
zero temperature as $\sim \sqrt{T/\Gamma}$. To explain that we can proceed
along the same line of reasoning as in the non-interacting case. It is
reasonable to assume that in this particular situation the dominant
contribution to $G$ is generated near the resonance, when
$\Delta$ is aligned with the chemical potentials in the leads. The
averaged conductance of the structure is then proportional to the ratio of the
effective level width $w(T)$ to the amplitude $\Delta_1$ of the oscillation 
$G \sim w(T)/\Delta_1$. As we know from the solution of the symmetric static
case at $g=1/2$, the level width vanishes as a square root of temperature 
\cite{ng,ourPRL}. That is why we find the same power--law for $G(T)$.
Note that this is very much different from the non-interacting
case where, of course, $w(T)$ does not depend on temperature. 
In order to obtain information about 
$G$ at intermediate frequencies one has to solve equation (\ref{inteqfull})
and extract the transmission coefficient numerically. The outcome of this
approach is shown in Fig. \ref{Ipicture1}. As can be seen from
this figure, there is an important qualitative correction to the adiabatic
approximation. Namely, the exact $G$ saturates as 
$T\rightarrow 0$ at a finite $\Omega$-dependent value:
$G(T=0)\sim \sqrt{\omega}$. 
The on-set of this saturation takes place at $T^* \sim \Omega$. At
temperatures higher than $T^*$ the system finds itself in the low frequency
limit and $G$ follows the square root power-law found in the static
approximation. On the contrary, below $T^*$ the constriction can be considered
as being in the high frequency limit, where the conductance is expected to
saturate. The resonant constellation $\Delta(t) \sim 0$ ceases to generate the
dominant contribution becoming too narrow in comparison to the off-resonant
positions of the dot level, when the effective width is known to saturate
\cite{ng,ourPRL}.     

\begin{figure}
\vspace*{0.0cm}
\includegraphics[scale=0.3]{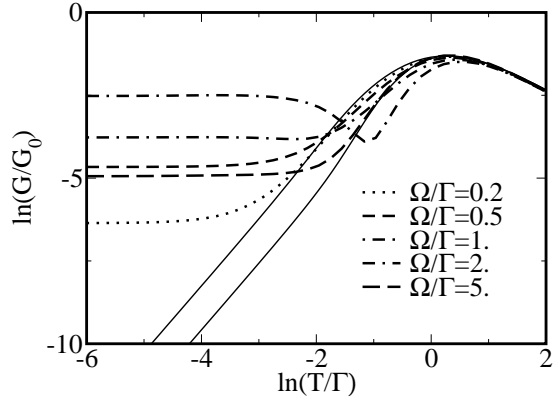}
\caption[]{\label{intfig2} Temperature dependence of the averaged
conductance in the interacting case for different gate frequencies $\Omega$
and $\Delta_0/\Gamma=2.$, $\Delta_1/\Gamma=1.$  . 
The upper solid line represents the result of the adiabatic approximation while
the lower one corresponds to the anti-adiabatic approximation.
Notice the double logarithmic scale.}
\vspace*{-0.0cm}
\end{figure}

The results for the `immersed' level, when $\Delta_0 < \Delta_1$ are very
similar, see Fig. \ref{intfig2}. Again, the conductance curves saturate at
temperatures $T \ll \Omega$. However, the (anti-)adiabatic approximation 
as well as the dynamic curves in the intermediate regime show  different
power-laws: instead the square root in the resonant ($\Delta_0=0$) situation
the conductance decays toward low temperatures quadratically. This power-law
originates 
in the different transport mechanism: now the most of the conductance does
not come from the resonant transmission, which cannot be achieved any more, 
but from the off-resonance processes which show quadratic temperature 
behaviour. Moreover, in the immersed case the adiabatic
approximation yields higher conductances than the anti-adiabatic ones. 
The asymptotic conductance 
behaviour at $T=0$ shows another interesting feature. 
It is maximal when the oscillation
frequency matches the off-set energy $\Delta_0$. In that situation the
tunnelling electron can absorb an energy quantum from the gate and get
transmitted through the system as if there is no energy off-set.  
 
\section{Conclusions}
We have analysed transport properties of a single site quantum dot coupled to
two LL electrodes in the case of an AC modulated gate. 
Using a unitary transformation and refermionization  we were
able to recast the Hamiltonian into a quadratic form at a special
interaction strength $g=1/2$. This allowed for an exact calculation of the
experimentally relevant period averaged conductance. Its temperature dependence
turns out to possess a rich behaviour, showing two cross--over points. 
Upon lowering the temperature for $T\gg \Gamma$ the effect of the AC
modulation is not visible and $G(T)$ follows the same power-law as in the 
static case $\sim \Gamma/T$ no matter how large is the off-set energy 
$\Delta_0$. In the case of $\Delta_0=0$ and for oscillation frequencies 
$\Omega \gg \Gamma$ the averaged conductance saturates  at a value 
larger than $G(T=\Gamma)$. However, as soon as $\Omega$ becomes smaller than 
$\Gamma$, the conductance vanishes first as a square root of temperature
in the interval between $T=\Omega$ and  $T=\Gamma$ and after that saturates 
at a finite value. The saturation toward low temperature is the most
prominent result of a dynamic gating and has a potential to obscure the
resonant tunnelling physics occurring in strictly static set-ups. Nevertheless,
an AC modulation experiment has additional advantages against the static 
one - it enables an extraction of the temperature dependence of the
resonance peak width from the information about $G(T)$ in the intermediate
temperature regime $\Omega<T<\Gamma$.

Our findings are, strictly speaking, restricted to the spinless $g=1/2$ case. 
Based on this results and the achieved understanding of 
various conductance mechanisms, we can however speculate about the
power-laws that will emerge in the general-$g$ case  
as well as in set-ups involving electrons with spin
(so long as Kondo physics is not important) and SWNTs. 
For the spinless system the most important regimes are the following: (i) the
high temperature limit $G(T) \sim T^{2g-2}$; (ii) the intermediate regime $G(T)
\sim T^{1-g}$; (iii) saturation to a frequency dependent value
at lowest temperatures. 
The results for the resonant tunnelling between two spinful LLs and SWNTs are
obtained by a substitution $g\rightarrow (g+1)/2$ for a system with spin and 
$g\rightarrow (K+3)/4$ for an SWNT, where $K$ is the nanotube 
Luttinger parameter. We shall present the detailed analysis of the
general-$g$ case in a separate publication \cite{generalg}. 

To summarise, the dynamic resonant conductance for LL leads is
markedly different from that for non-interacting contacts. It
possesses various distinctive features which are, in principle,
verifiable experimentally. In our view a comparison
of static and dynamic (AC-modulated) conductance measurements on
the same system would be especially valuable.


\appendix
\section{Adiabatic approximation}\label{AppendixA}
In this Appendix, we elaborate on the exact formula (\ref{exNIresult})
for the conductance in the non-interacting case at $T=0$. 
Expanding (\ref{exNIresult}) in $T$ we obtain:
\begin{equation}\label{G00}
G/G_0 =\sum\limits_{n=-\infty}^\infty\frac{J_n^2(\Delta_1/\Omega)}{
(n\Omega-\Delta_0)^2+1}\;. 
\end{equation}

First of all, it is useful to notice that  $G/G_0$, as
given by formula (\ref{G00}), is always smaller than $1$.
This immediately follows from the well known identity 
(see, e.g., \cite{watson})
\[
\sum\limits_{n=-\infty}^\infty J_n^2(z)= 1\;.
\]

The anti-adiabatic case ($\Omega \gg \Delta_1$) is trivial. For $\Delta_0=0$
the perfect conductance is recovered. Expanding the Bessel
functions, we find the next--to--leading correction as
\[
G/G_0=1-\frac{1}{2}\frac{\Omega^2}{\Omega^2+1}
\left(\frac{\Delta_1}{\Omega}\right)^2+...
\]
Note that the perturbative correction in $\Delta_1$ is
converging for all $\Omega$.

Next we consider the adiabatic case $z= \Omega/\Delta_1\ll 1$.
This turns out to be more involved. Using another well known
identity (see \cite{watson}), 
\[
\sum\limits_{n=-\infty}^\infty J_n^2(z)e^{-i n\phi}=
J_0[2z\sin(\phi/2)]\;,
\]
we re-write formula (\ref{G00}) as

\begin{equation}\label{Gint}
G/G_0=\frac{1}{\pi}\mbox{Re}
\int\limits_{0}^\pi d\phi
J_0[2z\sin(\phi/2)] f(\phi)\;,
\end{equation}
with 
\[
f(\phi)=\sum\limits_{n=-\infty}^\infty  \frac{e^{i n\phi}}
{(n\Omega-\Delta_0)^2+1}\;.
\]
This sum can be evaluated by using Poisson
summation formula with the result
\[
f(\phi)=\frac{\pi}{\Omega}
\frac{e^{(\pi-\phi)(1-i\Delta_0)/\Omega}}{2\sinh[\pi(1-i\Delta_0)/\Omega]}
+\frac{\pi}{\Omega}
\frac{e^{(\phi-\pi)(1+i\Delta_0)/\Omega}}{2\sinh[\pi(1+i\Delta_0)/\Omega]}
\;.
\]
When substituting the above expression into formula (\ref{Gint}) 
for conductance we observe that, in the $\Omega\to 0$ limit, 
the $\phi$-integral converges in a narrow interval $\phi\sim \Omega$. 
Rescaling $\phi=\Omega\theta$ and taking $\Omega\to 0$ limit we find
\begin{equation}\label{Gstaticb}
\lim\limits_{\Omega\to 0}G/G_0= \mbox{Re}
\int\limits_{0}^\infty d\theta e^{-(1-i\Delta_0)\theta} 
J_0(\Delta_1\theta)
={\rm Re} \frac{1}{\sqrt{(1-i\Delta_0)^2+\Delta_1^2}}\;.
\end{equation}
This result justifies the use of the adiabatic approximation
adopted in the main text.

\end{document}